# Experimental Demonstration of Five-photon Entanglement and Open-destination Teleportation

Zhi Zhao[1], Yu-Ao Chen[1], An-Ning Zhang[1], Tao Yang[1], Hans Briegel[2] & Jian-Wei Pan[1,3]

[1] Department of Modern Physics, University of Science and Technology of China, Hefei, Anhui, 230027, PR China

[2] Sektion Physik, Ludwig-Maximilians-Universität München, Theresienstrasse 37, D-80333 München, Germany

[3] Physikalishes Institut, Universitaet Heidelberg, Philosophenweg 12, D-69120 Heidelberg, Germany

**Universal quantum error-correction requires the ability of manipulating entanglement of five or more particles[1,2]. Although entanglement of three[3,4] or four[5,6] particles has been experimentally demonstrated and used to obtain the extreme contradiction between quantum mechanics and local realism[7,8], the realization of five-particle entanglement remains an experimental challenge. Meanwhile, a crucial experimental challenge in multi-party quantum communication and computation is the so-called open-destination teleportation[9,10]. During open-destination teleportation, an unknown quantum state of a single particle is first teleported onto a $N$-particle coherent superposition to perform distributed quantum information processing[11,12]. At a later stage this teleported state can be readout at any of the $N$ particles for further applications by performing a projection measurement on the remaining $N$-1 particles. Here, we report a proof-of-principle demonstration of five-photon entanglement and open-destination teleportation. In the experiment, we use two entangled photon pairs to generate a four-photon entangled state, which is then combined with a single photon state to achieve the experimental goals. The methods developed in our experiment would**

have various applications e.g. in quantum secret sharing [13,14] and measurement-based quantum computation[11,12].

In our experiment, the way to entangle five photons is a straightforward generalization of the schemes suggested for observation of three- and four-photon Greenberger-Horne-Zeilinger (GHZ) entanglement[15,16]. As shown in Fig.1a, we start from two polarization-entangled photon pairs 2-3 and 4-5, which are in the state $|\Phi^+\rangle_{ij}$. We use the usual Bell states

$$|\Phi^\pm\rangle_{ij} = \frac{1}{\sqrt{2}}(|H\rangle_i|H\rangle_j \pm |V\rangle_i|V\rangle_j)$$
$$|\Psi^\pm\rangle_{ij} = \frac{1}{\sqrt{2}}(|H\rangle_i|V\rangle_j \pm |V\rangle_i|H\rangle_j)$$
, (1)

where $H$ and $V$ denote horizontal and vertical linear polarizations, and $i$ and $j$ index the spatial modes of the photons. One photon out of each pair (3 and 4) is then steered to a polarization beam splitter (PBS) where the path lengths of each photon have been adjusted such that they arrive simultaneously. Since the PBS transmits $H$ and reflects $V$ polarization, coincidence detection between the two outputs of $PBS_{34}$ implies that either both photons 3 and 4 are $H$-polarized or both $V$-polarized, and thus projects the four-photon state onto a two-dimensional subspace spanned by $|H\rangle_2|H\rangle_3|H\rangle_4|H\rangle_5$ and $|V\rangle_2|V\rangle_3|V\rangle_4|V\rangle_5$. After the $PBS_{34}$, the renormalized state corresponding to a four-fold coincidence (before the $PBS_{12}$) is

$$|\Phi\rangle_{2345} = \frac{1}{\sqrt{2}}(|H\rangle_2|H\rangle_3|H\rangle_4|H\rangle_5 + |V\rangle_2|V\rangle_3|V\rangle_4|V\rangle_5),$$ (2)

which exhibits four-photon GHZ entanglement[6,15].

To generate five-photon GHZ entanglement, we further prepare photon 1 in the state $\frac{1}{\sqrt{2}}(|H\rangle_1 + |V\rangle_1)$ and adjust the path lengths of photons 1 and 2 such that they also arrive at the $PBS_{12}$ simultaneously. Then, it is easy to see that after the photons pass through the two PBS, the state corresponding to a five-fold coincidence is given by

$$|\Phi\rangle_{12345} = \frac{1}{\sqrt{2}}(|H\rangle_1|H\rangle_2|H\rangle_3|H\rangle_4|H\rangle_5 + |V\rangle_1|V\rangle_2|V\rangle_3|V\rangle_4|V\rangle_5).$$ (3)

This is exactly a five-photon GHZ state.

The scheme described above deserves some further comments. First, the five-photon entanglement here is observed only under the condition that there is one and only one photon in each of the five output modes. This conditional feature, however, does not prevent us from performing an experimental test of quantum non-locality[7,8] or practical applications in linear optics quantum information processing (QIP) [17,18].

Moreover, the setup sketched in Fig. 1a can also be used to realize open-destination teleportation, or equivalently, a sort of encoding-decoding operation. In the open-destination scheme, we are going to use the four-particle GHZ state (2) as the resource and photon 1 as the "teleportee". To see this, let us consider a specific example (Fig. 1b). Suppose that photon 1 is in a general unknown polarization state $|\Psi\rangle_1 = \alpha|H\rangle_1 + \beta|V\rangle_1$. In order to overcome the unavoidable decoherence effect on the state $|\Psi\rangle_1$, one could exploit quantum error correction. For example, to detect and correct the bit-flip error[2,19], we could encode the unknown state of photon 1 into a three-qubit superposition, say $|\Psi\rangle_{345} = \alpha|H\rangle_3|H\rangle_4|H\rangle_5 + \beta|V\rangle_3|V\rangle_4|V\rangle_5$. Note that, to realize universal quantum error correction, a more complicated encoding procedure is needed.

To achieve the three-qubit encoding, we first prepare photons 2, 3, 4 and 5 in the four-photon GHZ state $|\Phi\rangle_{2345}$ of equation (2). Then, using the four Bell states of photons 1 and 2 the overall state of the five photons can be rewritten as:

$$\begin{aligned}|\Psi\rangle_{12345} &= |\Psi\rangle_1|\Phi\rangle_{2345}\\ &= \frac{1}{2}[|\Phi^+\rangle_{12}(\alpha|H\rangle_3|H\rangle_4|H\rangle_5 + \beta|V\rangle_3|V\rangle_4|V\rangle_5)\\ &+ |\Phi^-\rangle_{12}(\alpha|H\rangle_3|H\rangle_4|H\rangle_5 - \beta|V\rangle_3|V\rangle_4|V\rangle_5)\\ &+ |\Psi^+\rangle_{12}(\alpha|V\rangle_3|V\rangle_4|V\rangle_5 + \beta|H\rangle_3|H\rangle_4|H\rangle_5)\\ &+ |\Psi^-\rangle_{12}(\alpha|V\rangle_3|V\rangle_4|V\rangle_5 - \beta|H\rangle_3|H\rangle_4|H\rangle_5)] \end{aligned} \quad (4)$$

This implies that a joint Bell measurement on photons 1 and 2 would thus project the state of photons 3, 4 and 5 into one of the four corresponding states as shown in equation (4).

Depending on the measurement results on photons 1 and 2, one can then perform a local unitary transformation, independent of $|\Psi\rangle_1$, on photons 3, 4 and 5 to convert the state into a three-particle superposition of $|\Psi\rangle_{345}$. This is how encoding works. During encoding operation, the unknown state of a single particle is teleported onto a three-particle superposition and can thus be protected against bit-flip error[2,19].

Obviously, to demonstrate the working principle of encoding operation it is sufficient to identify one of the four Bell-states (say $|\Phi^+\rangle_{12}$) although this would result in a reduced efficiency – the fraction of success – of 25%. The encoding operation can be achieved using the setup shown in Fig. 1a. First, the setup provides the necessary four-photon GHZ entanglement $|\Phi\rangle_{2345}$ after the photons 3 and 4 passes through the PBS$_{34}$. Second, as demonstrated in a recent experiment[6], the projection onto state $|\Phi^+\rangle_{12}$ can be accomplished by performing a joint polarization measurement in the $+/-$ basis behind the PBS$_{12}$, where $|\pm\rangle = \frac{1}{\sqrt{2}}(|H\rangle \pm |V\rangle)$. As stated in ref. 6, registering a $|+\rangle_1 |+\rangle_2$ or $|-\rangle_1 |-\rangle_2$ coincidence acts as a projection onto $|\Phi^+\rangle_{12}$. This projection leaves the photons 3, 4 and 5 in the state $|\Psi\rangle_{345} = \alpha |H\rangle_3 |H\rangle_4 |H\rangle_5 + \beta |V\rangle_3 |V\rangle_4 |V\rangle_5$.

At a later stage, we can then utilize decoding operation to read out or apply desired specific operations on the unknown state for further quantum information processing. To do so, we perform a local polarization measurement on two of the three photons 3, 4 and 5 in the $+/-$ basis. For example, if we perform a $\pm$ polarization analysis on photons 4 and 5, then if the measurement results on these two photons are the same – that is $|+\rangle_4 |+\rangle_5$ or $|-\rangle_4 |-\rangle_5$ – then photon 3 is left in the state $|\Psi\rangle_3 = \alpha |H\rangle_3 + \beta |V\rangle_3$. If the results are opposite, namely $|+\rangle_4 |-\rangle_5$ or $|-\rangle_4 |+\rangle_5$, then photon 3 is left in the state $\alpha |H\rangle_3 - \beta |V\rangle_3$. In the second case, one can simply perform a local phase flip operation on photon 3 to convert its state into $|\Psi\rangle_3$,

i.e. the original state of photon 1. We can thus teleport the unknown quantum state from photon 1 to photon 3.

With emphasis, we would like to note that, in a similar manner the initial state of photon 1 can also be teleported either onto photon 4 or photon 5 by performing a polarization measurement either on photons 3 and 5 or on photons 3 and 4 in the ± basis. In contrast to the original teleportation scheme[9], after the encoding operation the destination of teleportation is left open until we perform a polarization measurement on two of the remaining three photons. This implies that, even though photons 3, 4 and 5 are separated far away from each other, one can still choose which particle should act as the output where the initial state of photon 1 is transferred to[10]. This is why we have called such an encoding-decoding procedure as open-destination teleportation. It is therefore a generalization of standard teleportation, when no prior agreement on the final destination of the teleportation is necessary. Open-destination teleportation is possible if the parties initially share a multi-party entangled state, such as the GHZ state, but more generally any graph state[21] (of which the cluster state[20] is another example) can be used. Graph states are natural generalizations of the EPR state, which play a key role in novel schemes of quantum information processing[21]. A potential application of the open-destination teleportation is the scheme of quantum computation introduced by Gottesman and Chuang[11] using GHZ states and teleportation as a resource to realize quantum gates. Similarly, the ability to establish flexible teleportation channels with an entangled (cluster) state is a central ingredient in the one-way quantum computer[12].

Although our scheme to manipulate five-photon entanglement is theoretically simple, its experimental realization is very challenging. So far, spontaneous parametric down-conversion (SPDC) is still the best source of entangled photons, which is therefore to be used as the basic entanglement resource in the present experiment. However, owing to the probabilistic feature of SPDC the coincidence count rate in our five-photon experiment would be very low. To overcome this difficulty, we are going to combine a high intensity source of entangled photons and a source of pseudo-single photons so that we can collect enough experimental data to confirm the existence of five-photon entanglement within a reasonable time.

A schematic drawing of our experimental setup is shown in Fig. 2. In the experiment, a light pulse from a mode-locked Ti:Saphhire laser (with a duration of 200fs, a repetition rate of 76MHz and a central wavelength of 788nm) first passes through a frequency doubler, i.e. the LBO crystal (LiB O ). Behind the LBO, three dichroic beamsplitters (DM) are used to separate the mixed ultraviolet (UV) and infrared light components. The UV pulse further passes through a beta barium borate (BBO) crystal twice to generate two entangled photon pairs in the input modes 2-3 and 4-5[22], where both pairs are in the desired Bell-state $|\Phi^+\rangle$. The transmitted near-infrared pulse (part of the original laser beam) is attenuated by combining a ultra-fast laser output coupler mirror and two polarizers to a weak coherent state such that there is only a very small probability of containing a single photon for each pulse. We can thus prepare the required single photon state[16]. Note that, such a pseudo-single photon source has been used in numerous experiments[23,24]. Throughout the experiment, the coincidence time-window is set to be 4ns, which ensures the accidental coincidence is negligible.

In order to prepare a stable high intensity source of entangled photons, various efforts have been made. First, we properly focus the infrared pulse on the LBO crystal to achieve the best up-conversion efficiency. Meanwhile, to avoid the damage to the LBO crystal caused by the focusing laser beam we assemble the LBO crystal in a closed but transparent oxygen tube. Second, by using a compact set-up and by focusing the UV pump onto the BBO crystal we achieve both better collection efficiency and production rate of entangled photon pairs. Third, two CCD cameras C1 and C2 are used to monitor the laser beam direction. Thus, with the help of feedback loop we can well stabilize the beam direction. In this way, we finally managed to obtain an average UV pump power of 480mW and observe an average two-fold coincidence of $2.4 \times 10^4$/sec both in modes 2-3 and 4-5. Compared to the high intensity source of entangled photons developed in the recent experiments[25], our source is not only brighter, and most significantly, it is also a ultra-stable one - which can be easily stabilized for a couple of days.

In order to implement five-photon entanglement, we first overlap the two photons in the input modes 3 and 4 at the $PBS_{34}$ where the path length of the photon 4 has been adjusted by the Delay 1 such that both photons arrive simultaneously. To make photons 3 and 4 indistinguishable, we have to guarantee that the two photons have perfect spatial and temporal overlap at the $PBS_{34}$. To achieve this, the two outputs of the $PBS_{34}$ are spectrally filtered ($\Delta\lambda_{FWHM} = 3nm$) and monitored by fibre-coupled single photon detectors[26]. Thus, according to ref. 15, the fourfold coincidence among the detectors D2, D3, D4 and D5 corresponds to the four-photon GHZ state of equation (2). The coherent superposition between the terms $|H\rangle_2|H\rangle_3|H\rangle_4|H\rangle_5$ and $|V\rangle_2|V\rangle_3|V\rangle_4|V\rangle_5$ was confirmed by observing an excellent interference visibility of 82% in the $+/-$ basis (see Fig.3 a).

To further generate five-photon entanglement, we then add another $PBS_{12}$ and vary the path length of the single photon by the Delay 2 such that photons 1 and 2 arrive at the $PBS_{12}$ simultaneously. In a similar way, we can make photons 1 and 2 indistinguishable by using spectral filtering and fibre-coupled photon detectors. To ensure that the two photons in the input modes 1 and 2 have a good spatial and temporal overlap at the $PBS_{12}$, we can descend the $PBS_{34}$ to observe the visibility of three-photon entanglement where the threefold coincidence among the detectors D1, D2, and D3 is a conditional three-photon GHZ entanglement[16]. In the experiment, the intensity of single photon source, i.e. the average photon number per pulse, is attenuated to about 0.05. By combining the single photon source with the entangled pair of photons 2 and 3, a three-fold coincidence of 500 per second is obtained, which is three orders of magnitude higher than a recent experiment[27]. Again, the best visibility will be obtained at zero delay. In our three-photon entanglement, the visibility at zero delay was observed to be 68% (see Fig. 3b).

After achieving the perfect time overlap between photons 1 and 2, the PBS$_{34}$ was ascended back to its old position. According to equation (3), the five-fold coincidence among the detectors D1, D2, D3, D4 and D5 will then exhibit five-photon GHZ entanglement.

To experimentally verify that the five-photon entanglement has been successfully obtained, we first show that under the condition of registering a five-fold coincidence only the $|H\rangle|H\rangle|H\rangle|H\rangle|H\rangle$ and $|V\rangle|V\rangle|V\rangle|V\rangle|V\rangle$ components are observed, but no others. This was done by comparing the counts of all thirty-two possible polarization combinations, $|H\rangle|H\rangle|H\rangle|H\rangle|H\rangle$, ..., $|V\rangle|V\rangle|V\rangle|V\rangle|V\rangle$. The measurement results (Fig. 4a) in the $H/V$ basis show that the signal-to-noise ratio, defined as the ratio of any of the desired components to any of the thirty other non-desired ones, is about 40:1 on average.

Second, we further perform a polarization measurement in the $+/-$ basis in order to verify that the two terms of $|H\rangle|H\rangle|H\rangle|H\rangle|H\rangle$ and $|V\rangle|V\rangle|V\rangle|V\rangle|V\rangle$ are indeed in a coherent superposition. Transforming $|\phi\rangle_{12345}$ to the $+/-$ linear polarization basis yields an expression containing 16 (out of 32 possible) terms, each with an odd number of $|+\rangle$ components. Combinations with even numbers of $|+\rangle$ components do not occur. As a test for coherence we can check the presence or absence of various components. In Fig. 4b we compare the $|+\rangle|+\rangle|+\rangle|+\rangle|+\rangle$ and $|+\rangle|+\rangle|+\rangle|+\rangle|-\rangle$ count rates as a function of the pump delay 1 mirror position while the delay 2 is at zero delay. When the delay 1 is also at zero delay, the unwanted terms were found to be suppressed with an average visibility of $0.59 \pm 0.07$, which is sufficient to violate a five-particle Bell-type inequality imposed by local realism[28]. Therefore, the measurement results clearly demonstrate that for the first time the five-particle entanglement has been observed.

We now show how the same setup can be used to implement the open-destination teleportation. To demonstrate that our open-destination teleportation protocol works for a general unknown polarization state of photon 1, we choose to teleport $|+\rangle$ and $|-\rangle$ linear polarization states $\frac{1}{\sqrt{2}}(|H\rangle \pm |V\rangle)$ and right-hand ($|R\rangle$) and left-hand ($|L\rangle$) circular polarization states $\frac{1}{\sqrt{2}}(|H\rangle \pm i|V\rangle)$. In the experiment, we decided to analyze the Bell-state $|\Phi^+\rangle_{12}$. The required projection onto $|\Phi^+\rangle_{12}$ was achieved by registering a $|+\rangle_1|+\rangle_2$ coincidence behind the PBS$_{12}$. Then, as we already discussed, conditional on a $|+\rangle|+\rangle$ coincidence detection in the output modes 3 and 4, photon 5 will be left in the initial state of photon 1, that is, the unknown state of photon 1 is teleported to photon 5. In order to demonstrate the initial state of photon 1 can also be teleported to some other location, for example location 4, we can, in a similar manner, perform a polarization analysis on the photons 3 and 5 in the $+/-$ linear polarization basis. Then conditional on a $|+\rangle|+\rangle$ coincidence detection in the output modes 3 and 5, photon 4 will be left in the initial state of photon 1.

In our experiment, the integration time for each teleportation measurement is about 10 hours. As shown in Fig. 4b, in every polarization analysis measurement the five-fold coincidence rate is about 100 for the maximum (desired) component and 20 for the minimum (non-desired) component per 10 hours. The experimental fidelities of teleportation from photon 1 to photon 5 and from photon 1 to photon 4 for $+/-$ linear and $R/L$ circular polarization states are shown in Table 1. From Table 1, one can see that all the observed teleportation fidelities ($\sim 0.80 \pm 0.04$) are well above the classical limit of two-thirds, hence fully demonstrating open-destination teleportation.

Although compared to the previous experiments[3,6,29] our experimental demonstration of five-photon entanglement and open-destination teleportation might seem to be only a modest step forward, the implications are rather profound. First, our experiment demonstrated for the first time the ability to manipulate five-particle entanglement, which is the threshold number of qubits required for universal error-correction[1,2]. Second, the realization of open-destination teleportation opens up new possibilities for distributed QIP[11,12]. Finally, the techniques developed in the present experiment enables experimental investigations on a number of quantum protocols. While our high intensity three-photon entanglement source immediately allows an experimental realization of third-man quantum cryptography[3] and quantum secret sharing[13,14], the five-photon experimental setup would also allow an experimental demonstration of both bit-flip error rejection for quantum communication[19] and a non-destructive CNOT gate for quantum computation[30].

**Acknowledgements**

This work was supported by the National Natural Science Foundation of China, Chinese Academy of Sciences and the National Fundamental Research Program (under Grant No. 2001CB309303 ).

Correspondence and requests for materials should be addressed to J.-W. Pan (jian-wei.pan@physi.uni-heidelberg.de).

**Table 1.** Fidelities of open-destination teleportation at two different locations

| polarization | Fidelities | |
| --- | --- | --- |
| | Location 4 | Location 5 |
| $|+\rangle$ | $0.79 \pm 0.04$ | $0.80 \pm 0.03$ |
| $|-\rangle$ | $0.81 \pm 0.04$ | $0.77 \pm 0.04$ |
| $|R\rangle$ | $0.75 \pm 0.04$ | $0.79 \pm 0.04$ |
| $|L\rangle$ | $0.79 \pm 0.04$ | $0.82 \pm 0.03$ |

**Figure Captions**

**Figure 1** Diagrams showing the principles of generating five-photon entanglement and of achieving open-destination teleportation. **a**, Two Einetsin-Podolsky-Rosen (EPR) entangled photon sources each emits one polarization-entangled photon pair, which is in the state $|\Phi^+\rangle = 1/\sqrt{2}(|H\rangle|H\rangle + |V\rangle|V\rangle)$. The single photon source (S) emits a single photon state $\frac{1}{\sqrt{2}}(|H\rangle_1 + |V\rangle_1)$. After the photons 1, 2, 3 and 4 pass through the two polarizing beam splitters $PBS_{12}$ and $PBS_{34}$, conditional on detecting one photon in each of the five output modes the five photons will exhibit five-photon Greenberger-Horne-Zeilinger (GHZ) entanglement. **b**, The GHZ entanglement source emits a four-photon maximally entangled state $|\Phi\rangle_{2345} = \frac{1}{\sqrt{2}}(|H\rangle_2|H\rangle_3|H\rangle_4|H\rangle_5 + |V\rangle_2|V\rangle_3|V\rangle_4|V\rangle_5)$. Photon 1 is in a general unknown polarization state $|\Psi\rangle_1 = \alpha|H\rangle_1 + \beta|V\rangle_1$. By performing a joint Bell-state measurement (BSM) between photons 1 and 2 we can follow the teleportation protocol to transfer the initial state of photon 1 to a multi-particle superposition $|\Psi\rangle_{345} = \alpha|H\rangle_3|H\rangle_4|H\rangle_5 + \beta|V\rangle_3|V\rangle_4|V\rangle_5$. Moreover, by further performing a polarization analysis on any two of the three photons 3, 4 and 5 in the $+/-$ basis one can convert the remaining photon into the initial state of photon 1.

**Figure 2** Set-up for experimental demonstration of five-photon entanglement and open-destination teleportation. A infrared light pulse from a Mode-locked Ti:Sapphire laser passes through the LBO crystal to generate the UV pulse necessary for parametric down-conversion. After the mixed UV and infrared components being separated by the DM, the UV pulse passes through the BBO crystal twice to generate two polarization-entangled photon pairs 2-3 and 4-5 and the remaining infrared pulse is sent through a beam attenuator (Atten) and a polarizing beam splitter (PBS) to prepare the single photon state necessary for both five-photon entanglement and for open-destination teleportation. The mirrors Delay 1 and Delay 2 are used to adjust the time delay of the photons 3 and 4 and of the photons 1 and 2, respectively. In order to interfere the photons 1 and 2 at the $PBS_{12}$ and the photons 3 and 4 and the $PBS_{34}$, narrow band-width filters F ( $\Delta\lambda_{FWHM} = 3nm$ ) and fibre-coupled detectors D1, …, D5 have been used to ensure good temporal and spatial overlap of the photons 1 and 2 and of the photons 3 and 4. The half-wave plate $\lambda/2$ and the quarter-wave plate $\lambda/4$ in front of the $PBS_{12}$ are used to transform the horizontal polarization into linear $+/-$ or circular $R/L$ polarization. The polarizers P1,…, P5 oriented at $+/-$ basis and $\lambda/4$ plate in front of the detectors allow measurement of linear $+/-$ or circular $R/L$ polarization.

**Figure 3** Experimental results showing the procedure to achieve perfect temporal overlap for the photons 1 and 2, and for the photons 3 and 4. **a**, The two four-fold coincidence curves were obtained for the polarization settings of $|+\rangle_2 |+\rangle_3 |+\rangle_4 |+\rangle_5$ and $|+\rangle_2 |+\rangle_3 |+\rangle_4 |-\rangle_5$ as a function of the Delay 1 position (without inserting the $PBS_{12}$). To meet the condition of temporal overlap for the photons 3 and 4, we move the Delay 1 in small step to search for the position where the two photons have the same arrival time. Since maximum interference occurs at zero delay, by fixing the Delay 1 to the position where we observe the best four-photon interference visibility (i.e. the centre of the peak and dip) we can achieve perfect temporal overlap for the photon 3 and 4. **b**, The two three-fold coincidence curves were obtained for the polarization settings of $|+\rangle_1 |+\rangle_2 |+\rangle_3$ and $|+\rangle_1 |+\rangle_2 |-\rangle_3$ as a function of the Delay 2 position (without inserting the PBS$_{34}$). Similarly, by setting the Delay 2 position to the centre of the peak and dip of the three-fold coincidence we can achieve perfect temporal overlap for the photon 1 and 2.

**Figure 4** Experimental results for observation of five-photon entanglement. **a**, To verify a five-photon GHZ state has been successfully generated, we first measure the 32 possible components in the $H/V$ basis. The measurement results show that the signal-to-noise, i.e. the ratio of any of the desired components to any of the thirty other non-desired ones, is about 40:1 on the average. This well confirms that within the experimental precision only the desired *HHHHH* and *VVVVV* terms are present. **b**, To further demonstrate these two terms are indeed in a coherent superposition, we also perform a joint polarization measurement on all five photons in the $+/-$ basis. The two curves were obtained for the polarization settings of $|+\rangle_1 |+\rangle_2 |+\rangle_3 |+\rangle_4 |+\rangle_5$ and $|+\rangle_1 |+\rangle_2 |+\rangle_3 |+\rangle_4 |-\rangle_5$ as a function of the Delay 1 position while with the Delay 2 fixed. The difference between the two curves at zero delay confirms the five-photon GHZ entanglement.

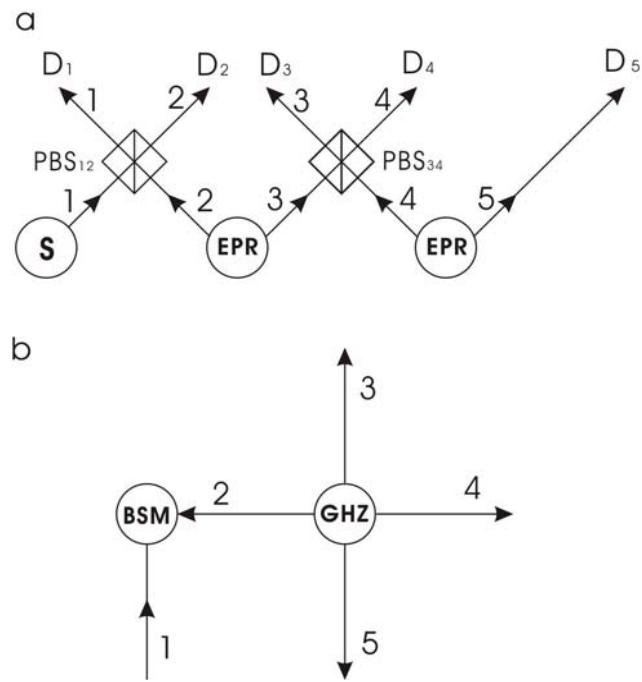

Figure 1

Figure 2

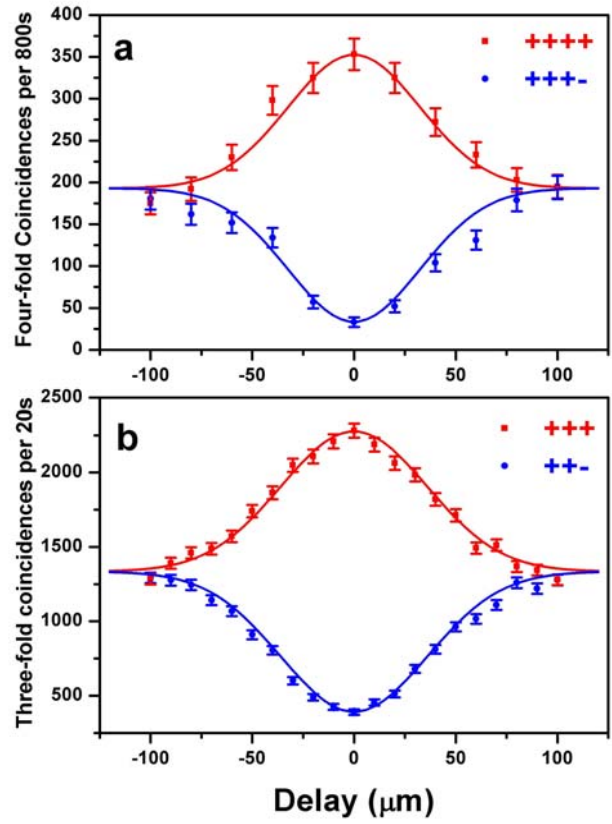

Figure 3

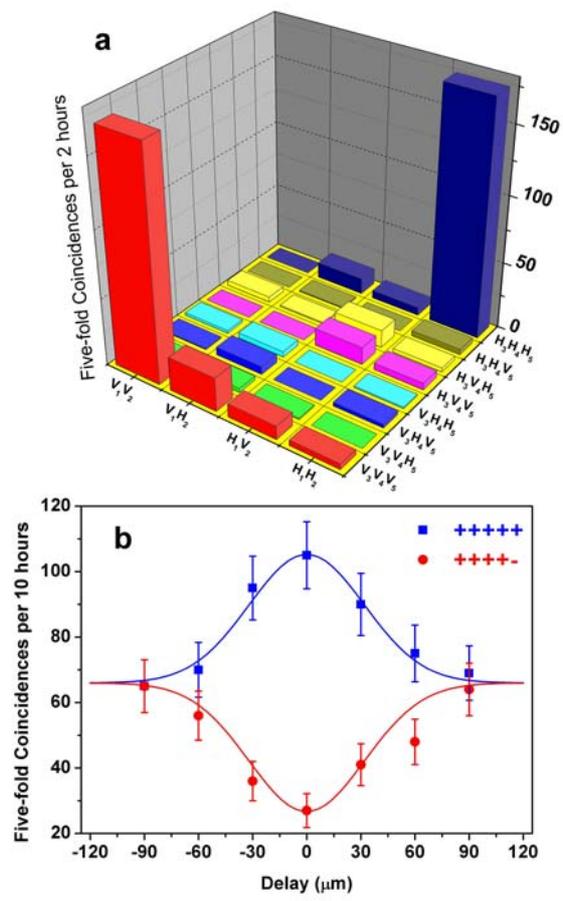

Figure 4